\documentclass[aps,prl,floatfix,twocolumn,superscriptaddress,showpacs]{revtex4}
\usepackage{amsmath}
\usepackage{amssymb}
\usepackage{epsfig}
\usepackage{bm}
\usepackage{graphics}
\begin{document}
\def\G{{\cal G}}
\def\F{{\cal F}}
\def\ea{\textit{et al.}}
\def\bM{{\bm M}}
\def\bN{{\bm N}}
\def\bV{{\bm V}}
\def\bj{\bm{j}}
\def\bSig{{\bm \Sigma}}
\def\bLam{{\bm \Lambda}}
\def\bfeta{{\bf \eta}}
\def\bc{{\bf c}}
\def\ba{{\bf a}}
\def\d{{\bf d}}
\def \xy{$x$--$y$ }
\def \xz{$x$--$z$ }
\def\bP{{\bf P}}
\def\bK{{\bf K}}
\def\bk{{\bf k}}
\def\bkn{{\bf k}_{0}}
\def\bx{{\bf x}}
\def\bz{{\bf z}}
\def\bR{{\bf R}}
\def\br{{\bf r}}
\def\bu{{\bm u}}
\def\bq{{\bf q}}
\def\bp{{\bf p}}
\def\bG{{\bf G}}
\def\bQ{{\bf Q}}
\def\bs{{\bf s}}
\def\bA{{\mathbf A}}
\def\bv{{\bf v}}
\def\b0{{\bf 0}}
\def\la{\langle}
\def\ra{\rangle}
\def\Im{\mathrm {Im}\;}
\def\Re{\mathrm {Re}\;}
\def\beq{\begin{equation}}
\def\eeq{\end{equation}}
\def\bea{\begin{eqnarray}}
\def\eea{\end{eqnarray}}
\def\bdm{\begin{displaymath}}
\def\edm{\end{displaymath}}
\def\bnab{{\bm \nabla}}
\def\Tr{{\mathrm{Tr}}}
\def\bJ{{\bf J}}
\def\bU{{\bf U}}
\def\bPsi{{\bm \Psi}}
\def\mA {\mathrm{A}}
\def\E {{\cal{E}}}
\def\brho{{\bm \rho}}

\title{Quantum phases of a two-dimensional dipolar Fermi gas}
\author{G.\ M.\ Bruun}
\affiliation{CNR-INFM BEC Center and Dipartimento~di~Fisica, Universit\`a di Trento, I-38050 Povo, Trento, Italy}
\affiliation{Niels Bohr Institute, University of Copenhagen, DK-2100 Copenhagen \O, Denmark}
\author{E.~Taylor}
\affiliation{CNR-INFM BEC Center and Dipartimento~di~Fisica, Universit\`a di Trento, I-38050 Povo, Trento, Italy}

\date{September 10, 2008}

\begin{abstract}
We examine the superfluid and collapse instabilities of a quasi two-dimensional gas of dipolar fermions aligned by an orientable external field.  
It is shown that the interplay between  the anisotropy of the  dipole-dipole interaction, the geometry 
of the system, and the $p$-wave symmetry of the superfluid order parameter means that the effective interaction for pairing can be made very large without the system collapsing. This leads 
to a broad region in the phase diagram where the system forms a 
stable superfluid. Analyzing the superfluid transition at finite temperatures, we calculate the Berezinskii--Kosterlitz--Thouless temperature as a function of the dipole angle.  
\end{abstract}


\maketitle

Trapped ultracold gases are increasingly being used to simulate solid-state systems, where clear experimental signatures of theoretical predictions are often lacking~\cite{Bloch}.  A limitation of these gases is that the interactions are typically $s$-wave~\cite{Reviews} whereas order parameters in solid state systems exhibit richer $p$- and $d$-wave symmetries.  Recent progress in the production of trapped, cold dipolar gases~\cite{Lahaye,Koch,Ospelkaus} promises to change this since the  dipole-dipole interaction is long-range and anisotropic.  
Importantly, the  interaction in fermionic heteronuclear Feshbach molecules 
can be large, with electric dipole moments on the order of a Debye~\cite{Kotochigova,Ospelkaus,Weber}.  This opens the door to 
experimentally reaching the superfluid phase. One complicating feature is that the gases tend to become unstable when interactions are sufficiently 
strong and attractive.  

\begin{figure}
\epsfig{file=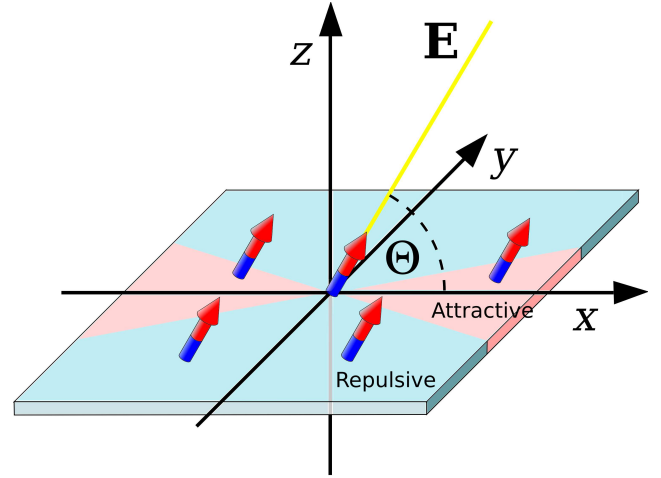, angle=0, width=0.47\textwidth}
\caption{In the proposed experimental setup, aligned Fermi dipoles are confined in the \xy plane.  The dipoles form an angle $\Theta$ with respect to the \xy plane. As the angle $\Theta$ is reduced, the region where the interaction between dipoles is attractive increases. }
\label{config}
\end{figure}

In this letter, we propose confining a gas of fermionic dipoles of mass $m$
  in the 2D plane by 
a harmonic trapping potential $V_z(z)=m\omega_z^2z^2/2$.  For $\hbar\omega_z\gg \epsilon_F$ where $\epsilon_F=k^2_F/2m$ is the Fermi energy, determined by the 2D density $n_{2D}=k_F^2/4\pi$,    the system is effectively 2D. The dipoles are aligned by an external field ${\mathbf{E}}$ (see Fig.\ \ref{config}), subtending an angle $\Theta$ with respect to the \xy plane.  The interaction between two dipole moments ${\mathbf{d}}$ separated by ${\mathbf{r}}$  is  given by 
 $
 V({\mathbf{r}})=D^2(1-3\cos^2 \theta_{rd})/r^3
 $.
  Here, $D^2=d^2/4\pi\epsilon_0$ for electric dipoles 
 and $ \theta_{rd}$ is the angle between ${\mathbf{r}}$ and ${\mathbf{d}}$. The strength of the interaction is parametrized by the dimensionless ratio $g=2D^2k_F^3/(3\pi\epsilon_F)$.  Since we consider identical fermions  at low temperatures $T$,
  additional short range interactions are suppressed. We show that  in this configuration, the effective interaction for pairing has $p$-wave
  symmetry and can be made large without the system collapsing. This makes the 2D system of dipoles a promising 
  candidate to study quantum phase transitions and pairing with unconventional symmetry.   

\begin{figure}
\begin{center}
\epsfig{file=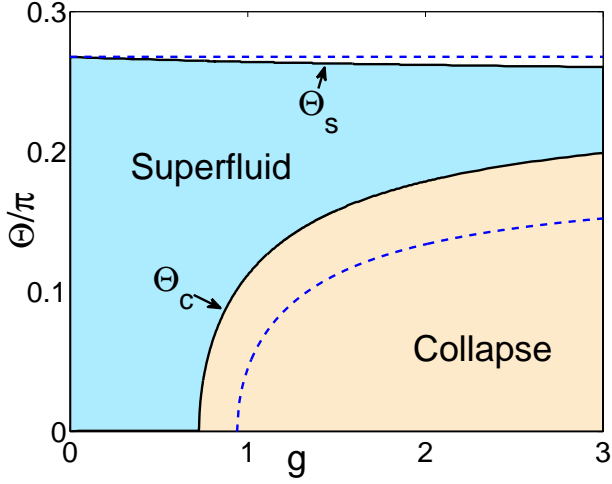, angle=0,width=0.47\textwidth}
\caption{The $T=0$ phase diagram. Phases are shown in terms of the interaction $g$.
For weak coupling, $g\lesssim0.725$, the Fermi pressure stabilizes the system for all angles. 
For stronger coupling, there is a critical angle $\Theta_c$ below which the attractive interaction overcomes the Fermi pressure and the gas collapses. 
 The system is superfluid and stable for $\Theta_c<\Theta<\Theta_s$.
The superfluid and normal regions are separated by a quantum phase transition.  
For comparison, we also plot as dashed lines the phase diagram calculated with no Fermi surface deformation effects ($\alpha=1$).}
\label{phase}
\end{center}
\end{figure}

A key point  is that the effective dipole-dipole interaction in the  \xy plane can be tuned by changing the angle $\Theta$.
This gives rise to several interesting effects. For $\Theta=\pi/2$, the interaction is repulsive and isotropic in the \xy plane and the dipoles are predicted to 
 undergo a quantum phase transition to a crystalline  phase for  
$g\simeq 27$ at $T=0$~\cite{Buchler}. As $\Theta$ is decreased, the interaction becomes anisotropic 
in the \xy plane and for $\Theta<\arccos(1/\sqrt{3})$, an \emph{attractive} sliver appears 
along the $x$ axis as illustrated in Fig.\ (\ref{config}).  This gives rise to two competing
phenomena: superfluidity and collapse~\cite{Koch}.  The resulting phase diagram is shown in Fig.\ (\ref{phase}) and is the main focus of this letter.

We first examine the critical angle $\Theta_c$ below which the attractive part of the interaction 
overcomes the Fermi pressure and the gas is unstable towards collapse. The instability is identified with a negative value of the inverse compressibility 
${\kappa}^{-1}=n_{2D}^2\partial^2 \E/\partial n^2_{2D}$. 
We analyze the stability of the dipolar Fermi gas using the normal phase energy density
 $\E_\mathrm{n}=\E_{\mathrm{kin}}+\E_{\mathrm{dir}}+\E_{\mathrm{ex}}$ with 
\bea 
\E_{\mathrm{kin}}= \frac{1}{(2\pi)^2}\int\! d^2\bk \frac{k^2}{2m} f_\bk= \frac{\pi}{2}\frac{n_{2D}^2}{m}(\frac{1}{\alpha^2}+\alpha^2),
\label{Ekin}
\eea
the kinetic energy density.  We allow for the possibility that the Fermi surface can deform~\cite{Miyakawa} due to the anisotropy of the interaction,
 by writing the $T=0$ Fermi distribution function as
$f_{\mathbf{k}}=\Theta(k_F^2-k_y^2/\alpha^2-k_x^2\alpha^2)$.
The direct and exchange energies are ($\br\equiv \br_1-\br_2$)
\bea
 \E_{\mathrm{dir}} = \frac{1}{2L^2}\int d^3\br_1d^3\br_2 n(\br_1,\br_1)V(\br)n(\br_2,\br_2),\label{Edir}
 \eea
and
\bea 
\E_{\mathrm{ex}} = -\frac{1}{2L^2}\int d^3\br_1d^3\br_2 n(\br_1,\br_2) V(\br)n(\br_2,\br_1)\label{Eex}
\eea
with  $n(\br_1,\br_2) \equiv \langle\hat{\psi}^{\dagger}(\br_1)\hat{\psi}(\br_2)\rangle$, $\hat{\psi}$ the dipolar field operator, 
 and $L^2$   the size of the gas.  In the superfluid phase, there is a contribution to the energy arising from the condensation energy $\E_\mathrm{c}\propto \Delta^2_0/\epsilon_F$
 which is small for $\Delta_0\ll \epsilon_F$.  We ignore this contribution in the following.

In the 2D limit $\hbar\omega_z\gg \epsilon_F$, all dipoles reside in the lowest harmonic oscillator level
$\Phi(z)$ in the $z$-direction. We then have 
 $n(\br_1,\br_2)=\Phi(z_1)\Phi(z_2)n_{\rm 2D}(\brho)$ with  
\begin{gather}
n_{\rm 2D}(\brho)=\int\frac{d^2k}{(2\pi)^2}f_{\mathbf{k}}e^{i{\mathbf{k\rho}}_{12}}
=\frac{k_F}{2\pi\tilde{r}({\mathbf{\rho}})}J_1[k_F\tilde{r}(\brho)].
\end{gather}
Here,  $J_1(x)$ is the Bessel function of first order, $\brho=(x,y)$, and 
$\tilde{r}(\brho)=\sqrt{x^2/\alpha^2+\alpha^2y^2}$. Evaluating the integrals in (\ref{Edir}) and (\ref{Eex}), we find
\beq
\E_{\mathrm{dir}}+\E_{\mathrm{ex}}=
-\frac{8n_{2D}^2}{3\pi m}gI(\alpha,\Theta)
\label{Einttight}
\eeq
where 
\beq
I(\alpha,\Theta)=\!\!\int_0^\infty\! \!\frac{dv}{v^2}\!\int_0^{2\pi}\!\!\!\! d\phi(1\!-\!3\cos^2\!\Theta\cos^2\!\phi)\!\Big[\frac{4J_1^2(t)}{t^2}-1\Big]
\label{I}
\eeq
is a dimensionless function of the Fermi surface deformation parameter $\alpha$ and the dipole angle $\Theta$, and $t\equiv v\sqrt{\alpha^2\sin^2\phi+\cos^2\phi/\alpha^2}$.
Equations (\ref{Ekin}), (\ref{Einttight}), and (\ref{I}) give the energy  as a function of  $g$, $\Theta$, and $\alpha$.
For a given coupling strength  $g$ and angle $\Theta$, $\alpha(g,\Theta)$ is found by minimizing  $\E$.  
In general, this has to be done  numerically. 
 However, for small Fermi surface deformation $\alpha \sim 1$ we can expand (\ref{I}) in the small parameter $t-v$. The resulting $\phi$ integrals 
are straightforward and using $\int_0^\infty dv[4J_1^2(v)/v^2-1]v^{-2}\simeq-0.9$ we obtain  
\begin{equation}
\alpha^4\simeq\frac{1+1.06g(1-\frac{9}{4}\cos^2\Theta)}{1+1.06g(1-\frac{3}{4}\cos^2\Theta)}.
\label{alphapert}
\end{equation}
Note that $\alpha<1$ as expected.

\begin{figure}[b]
\begin{center}
\epsfig{file=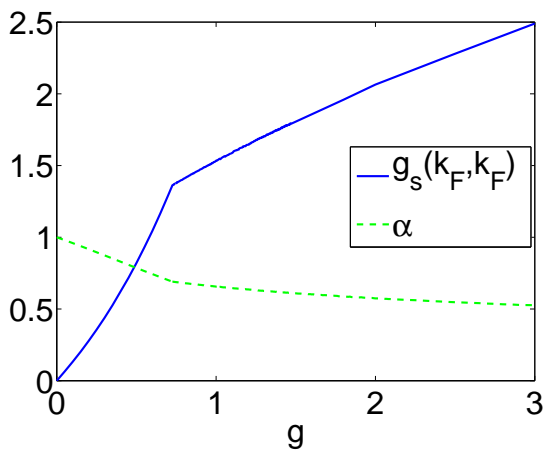, angle=0,width=0.4\textwidth}
\caption{ The deformation parameter $\alpha$ and the coupling strength $g_s(k_F,k_F)$ for $p$-wave pairing along the critical line for collapse $\Theta_c(g)$. 
For $g\lesssim0.725$ where there is no collapse, we have evaluated $\alpha$ and $g_s$ at $\Theta=0$.}
\label{GapFig}
\end{center}
\end{figure}

Since $\E\propto n^2_{2D}$, it is now straightforward to evaluate the compressibility $\kappa(g,\Theta)$.
The resulting region of the collapse instability is shown in Fig.\ \ref{phase}. 
There is a large region  where the system is stable even though the 
interaction is strong ($g>1$) and partly attractive [$\Theta<\arccos(1/\sqrt{3})$]. This is 
most easily understood by ignoring the Fermi surface deformation ($\alpha=1$), which yields
$\Theta_c\rightarrow\arcsin 1/\sqrt{3}$ for $g\rightarrow \infty$. In this limit, 
the interaction energy dominates and the gas collapses 
when the dipole-dipole interaction is attractive in more than half the \xy plane, i.e.\ for $\Theta<\arcsin 1/\sqrt{3}$.
Fermi surface deformation effects are significant however; in 
Fig.\  \ref{GapFig} we show how $\alpha(g,\Theta)$ deviates substantially
 from $1$ along the critical line $\Theta_c(g)$ for collapse.

The stability of the 2D system for a strong, attractive interaction 
makes it a promising candidate to observe superfluid pairing with unconventional symmetry. 
To investigate this, we solve the effective 2D BCS gap equation at $T=0$, derived from the usual 3D BCS pairing Hamiltonian using $\hat{\psi}(\br) = \hat{\psi}(\brho)\Phi(z)$.
  This ansatz results in a 2D gap equation in terms of an effective interaction $V_{2D}(\brho)= \int dz \Phi^2_{r}(z)V(\br)$, where $\Phi_{r}(z) = \exp(-z^2/2l^2_z)/\pi^{1/4}l^{1/2}_z$   with $l_z = \sqrt{2/m\omega_z}$
   is the lowest harmonic oscillator wavefunction for the reduced mass $m/2$. 
    Using the rescaled momentum $\tilde{\bk}=(\tilde k_x,\tilde k_y)=(\alpha k_x,k_y/\alpha)$ to describe pairing about
the deformed Fermi surface, the 2D gap equation becomes
\begin{equation}
\Delta_{\tilde{\mathbf{k}}}=-\int \frac{d^2\tilde\bk'}{(2\pi)^2}\tilde{V}_{2D}(\tilde{{\mathbf{k}}},{\tilde{\mathbf{k}}}')
\frac{\Delta_{\tilde{\mathbf{k}}'}}{2E_{\tilde{\mathbf{k}}'}}
\label{GapEqn}
\end{equation}
with $E_{\tilde\bk}=(\xi_{\tilde k}^2+\Delta_{\tilde\bk}^2)^{1/2}$, $\xi_{\tilde k}=\tilde k^2/2m-\mu$, and
$\mu$  the chemical potential. $\tilde{V}_{2D}(\tilde{{\mathbf{k}}},{\tilde{\mathbf{k}}}')$ is the Fourier transform
\beq
V_{2D}({\mathbf{k}},{\mathbf{k}}')=
\int d^2\brho\sin (\bk\cdot\brho) V_{\rm 2D}(\brho)\sin (\bk'\cdot\brho).
\label{Fourier}
\eeq
of $V_{2D}(\brho)$, expressed in terms of the scaled momenta $\tilde{\bk},\tilde{\bk}'$; i.e. $\tilde{V}_{2D}(\tilde{{\mathbf{k}}},{\tilde{\mathbf{k}}}')=V_{2D}({\mathbf{k}},{\mathbf{k}}')$. 
 Note that only $\sin$ components contribute to the Fourier series since the order parameter is antisymmetric.  The
Pauli exclusion principle therefore cancels the short range $r^{-3}$ 
divergence in the dipole-dipole interaction, making the Fourier transform (\ref{Fourier}) finite even in the 2D limit.

Using the expansions 
$\Delta_{\tilde{\mathbf{k}}}=\sum_{n=1}^\infty \Delta_{n}(\tilde{k})\cos[(2n-1)\phi]$ and  $\tilde{V}_{2D}(\tilde{\mathbf{k}},\tilde{\mathbf{k}}')=\sum_{nn'=1}^\infty V_{nn'}(\tilde k,\tilde k')\cos(2n-1)\phi\cos(2n'-1)\phi'$ in (\ref{GapEqn}),  
where $\phi$ is the angle between $\tilde{\mathbf{k}}$ and the $x$ axis, 
we find numerically that $V_{nn}\gg V_{nn'}$ for $n'\neq n$ and
all angles $0\le \Theta\le \pi/2$.  This shows that to a very good approximation the gap is $p$-wave, $\Delta_{\mathbf{k}}=\Delta_0\cos\phi$,
 and we can replace $\tilde{V}_{2D}(\tilde{\bk},\tilde{\bk}')$ with $V_{11}(\tilde{k},\tilde{k}')\cos\phi\cos\phi'$.  Using this in (\ref{GapEqn}), it reduces to
\begin{equation}
1=-\frac{4\pi}{m}\int \frac{d^2\tilde k'}{(2\pi)^2}\frac{g_s(\tilde k,\tilde k')\cos^2\phi'}{2E_{\tilde{\mathbf{k}}'}},
\label{GapEqn2}
\end{equation}
where $4\pi g_s/m \equiv V_{11}$.

The dimensionless effective pairing interaction $g_s$ corresponds to the dipole-dipole interaction averaged 
over the deformed Fermi surface, weighted by the $p$-wave symmetry of the order parameter.  After a straightforward but lengthy calculation, 
we find
\bea
g_s(k,k')&=&gM(k,k')I_s[\Theta,\alpha(g,\Theta)]\nonumber\\ &\simeq&  g M(k,k')(1-\frac{9}{4}\cos^2\Theta).
\label{PairInteraction}
\eea 
Here, $I_s$ and $M$ are dimensionless functions given by  
\begin{equation}
I_s=\int_0^{2\pi}\frac{d\phi}{2\pi}
\frac{(1+\cos2\phi)(1-\frac{3\cos^2\Theta\alpha^2\cos^2\phi}{\alpha^2\cos^2\phi+\sin^2\phi/\alpha^2})}{(\alpha^2\cos^2\phi+\sin^2\phi/\alpha^2)^{3/2}}
\label{Is}
\end{equation}
 and 
$
M(k,k')=\frac{k'^2}{2k_Fk}[(1+x)E(x)+(x-1)K(x)],
$
for $k<k'$  and $k'l_z\ll 1$; for $k'<k$ one simply swaps $k$ and $k'$.  $K(x)$ and $E(x)$ are the complete elliptic integrals of first and second 
kind respectively and $x=k^2/k'^2$. We have $M(k_F,k_F)=1$. For  $k'l_z\gtrsim 1$, one has $M(k,k')\propto k(k'l_z)^{-1}$ showing that
the 2D effective interaction has a high energy cut-off for $k'l_z\sim1$.
The last line in (\ref{PairInteraction}) is exact for $\alpha=1$.

We now analyze the $T=0$ gap equation (\ref{GapEqn2})  as the dipole angle $\Theta$ is adjusted.  Since $M(k,k')>0$ for all $k,k'$, 
the gap $\Delta_{\bk}$ is zero unless $I_s(\Theta,\alpha)<0$.  The resulting superfluid region found by evaluating the sign of $I_s$ 
is shown in Fig.\ \ref{phase}.  Numerically, we find that the Fermi surface deformation has little effect on the effective interaction for pairing as can be seen  in Fig.\ \ref{phase}. 
Ignoring Fermi surface deformation effects, we find from (\ref{PairInteraction}) that 
the system is superfluid  for $\Theta<\Theta_s=\arccos(2/3)$. The critical angle $\Theta_s$ for superfluidity is therefore to a good approximation independent of the interaction strength $g$ and is purely determined by geometry. 

Crucially, as $\Theta$ decreases from $\pi/2$, the system becomes unstable toward pairing  \emph{before} the gas collapses, i.e.\ $\Theta_s(g)>\Theta_c(g)$, 
and there is a significant region in the phase diagram where the system is superfluid yet stable.  Because the $p$-wave pair wavefunction is predominantly oriented along the $x$ axis 
where the interaction is maximally attractive, the effective pairing interaction (\ref{PairInteraction}) is stronger than the ``bare" 2D interaction in (\ref{Einttight}) that determines
 the collapse instability.  For this reason, even for strong interactions, there remains a window $\Theta_s<\Theta<\Theta_c$ where the system is a stable superfluid.

\begin{figure}
\begin{center}
\epsfig{file=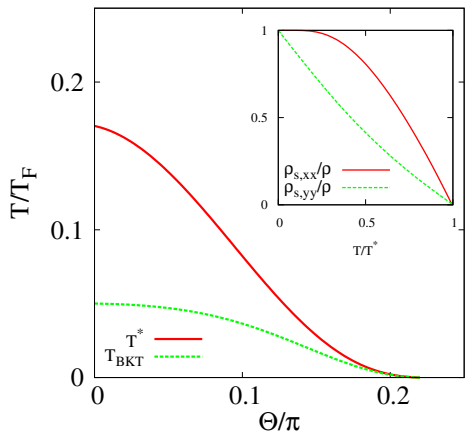, angle=0,width=0.47\textwidth}
\caption{
Berezinskii-Kosterlitz-Thouless ($T_{\mathrm{BKT}}$) and mean-field  ($T^*$) transition temperatures plotted as functions of $\Theta$ for $g=0.2$. Inset: Temperature-dependence of the diagonal components of the superfluid density for $\Theta=0$ and $g=0.2$. }
\label{TKTfig1}
\end{center}
\end{figure}

To examine the strength of the pairing interaction further, we make use of the fact that pairing occurs primarily at the Fermi surface, and in Fig.\  \ref{GapFig} we plot $g_s(k_F,k_F)$
from (\ref{PairInteraction}) along the critical line $\Theta_c(g)$ for collapse. This gives the largest possible attractive pairing interaction, before the system collapses.  
We see that the effective pairing interaction   increases monotonically with $g$ and can become very large.  Thus, one can produce a strongly paired gas without the system collapsing.
This should be compared with the 3D trapped case where recent results indicate that the system 
is superfluid and stable only in a  
 narrow region in phase space where the pairing is relatively weak~\cite{Miyakawa,Baranov04}. 
Solving the gap equation in the weak coupling regime yields
$\Delta_0(\Theta)=\epsilon_F4e^{-1/2}\sqrt{\Lambda}e^{1/g_s}$
for $g_s<0$ (i.e.\ $\Theta<\Theta_s$)
with $\Lambda=k^2_{\max}/k_F^2$ an ultraviolet cut-off. This shows that 
the superfluid phase transition is infinite order in the sense that $\partial_\Theta^n\Delta_0|_{\Theta_s}=0$ for all $n$.

At finite temperatures, long-range order is destroyed by phase fluctuations in 2D, and $\Delta_{\bk}(T>0)=0$.  The superfluid density remains finite, however, describing a Berezinskii-Kosterlitz-Thouless (BKT) superfluid of bound vortex-antivortex pairs~\cite{KTB}.  The critical temperature $T_{\mathrm{BKT}}$ for this phase is the temperature at which the free energy of a single unbound vortex vanishes~\cite{ChaikinLubensky},
\bea T_{\mathrm{BKT}} = \frac{\pi\hbar^2}{8m^2k_B}\bar{\rho}_s.\label{TKT0}\eea
Here,  $\bar{\rho}_s\equiv (\rho_{s,xx} + \rho_{s,yy})/2$ is the average of the diagonal components of the superfluid mass density tensor $\rho_{s,ij}$.  In estimating the energy $E = \frac{1}{2}\int d^2\br \rho_{s,ij}v_{i}v_{j}\approx \frac{\pi \hbar^2}{4m^2}\ln\left(\frac{L}{a}\right)\bar{\rho}_s$ of a single vortex of radius $a$ in a 2D box with sides of length $L$, we have assumed that the contribution from the off-diagonal component $\rho_{s,xy}v_{s,x}v_{s,y}$ due to the anisotropy of the vortex velocity field~\cite{Yi06} is small.  Generalizing the usual expression for $\rho_{s,ij}$~\cite{Leggett} to allow for the effects of Fermi surface deformation, one finds
\bea \rho_{s,ij} &=& mn\delta_{ij}-\frac{\beta}{4}\sum_{\tilde{\bk}}\mathrm{sech}^2\left(\frac{\beta E_{\tilde{\bk}}}{2}\right)\tilde{k}_{i}\tilde{k}_{j}.\label{rhos}\eea
As shown in the inset of Fig.~\ref{TKTfig1}, rotational symmetry is broken by the external field and $\rho_{s,xx}> \rho_{s,yy}$.  The $y$ component of the superfluid density tensor is suppressed since the quasiparticle spectrum is gapless in the direction perpendicular to the field. 

Since $\bar{\rho}_s$ depends on $T$, (\ref{TKT0}) has to be solved self-consistently. We determine 
$\bar{\rho}_s(T)$ from (\ref{rhos}) using a $T$-dependent gap $\Delta_{\bk}(T)$ 
calculated by including the usual Fermi functions in the gap equation.
The critical temperature 
determined this way is 
plotted in Fig.~(\ref{TKTfig1}) for $g=0.2$. We also plot the mean field transition temperature $T^*$ 
obtained from the gap equation. 
Phase fluctuations suppress $T_{\mathrm{BKT}}$ below  $T^*$. 
 For weak coupling, $T_{\mathrm{BKT}}\simeq T^*$.  Since $T^*\propto\sqrt{\epsilon_F\hbar\omega_z}$ and $n_{2D}\propto\epsilon_F$, 
we have $T_{\mathrm{BKT}}\propto n_{2D}^{1/2}$.  This suggests that one can cross the critical temperature 
by adiabatically expanding the gas keeping 
$T/\epsilon_F$ constant so that $T/T_{\mathrm{BKT}}\propto n_{2D}^{1/2}$ decreases~\cite{Petrov}.  For stronger interactions, $T_{\mathrm{BKT}}$ quickly approaches its upper bound 
$\frac{\pi\hbar^2}{2M^2k_B}\rho = T_F/16 \propto n_{2D}$.

Let us consider the experimental requirements for observing the effects discussed in this letter.
Recently, a gas of $^{40}$K-$^{87}$Rb
polar molecules was created with a density of $n\sim 10^{12}$cm$^{-3}$ at a temperature $T/T_F\gtrsim 2$~\cite{Ospelkaus}.
These molecules have dipole moments of $0.57$ Debye in their vibrational ground state.  Writing  $ g\simeq2\tilde d^2\tilde a^{-1}N_A/100$, where $\tilde d=d/1$Debye is the dipole moment in Debyes, $\tilde a=a/1\mu$m is the interparticle spacing measured in $\mu m$, and 
$N_A=m/m_u$ is the mass of the dipoles in atomic mass units, the experimental parameters of Ref.~\cite{Ospelkaus} give $g\simeq0.8$.
Also, the creation of 2D systems has already been achieved for bosons~\cite{Hadzibabic}.
 The observation of the effects discussed in this letter is therefore within experimental reach once further cooling 
has been achieved.

In conclusion, we studied the quantum phases of a dipolar Fermi gas in 2D aligned by an external field.    We demonstrated
 that by partially orienting the dipoles into the 2D plane, they can experience a strong $p$-wave pairing attraction without the system collapsing.  This makes the system a promising candidate to study quantum phase transitions and pairing with unconventional symmetry.  We also analyzed the BKT transition to the normal  state at non-zero $T$.

Stimulating discussions with S.\ Stringari are acknowledged.

\end{document}